\documentclass[]{article}

\usepackage{algorithmic}

\usepackage[english]{babel}

\newtheorem{e-proposition}[theorem]{Proposition}

\newtheorem{e-definition}[theorem]{Definition\rm}

\newcommand{\dis}{\displaystyle}
\newcommand{\OO}{{\mathcal O}}

\newcommand{\R}{{\mathbb R}}

\newcommand{\N}{{\mathbb{N} }}

\newcommand{\defo}{{ \nabla}}
\usepackage[]{algorithm2e}

\usepackage{amsmath}

\usepackage{color}

\usepackage{ulem}

\usepackage{nomencl}

\usepackage{amssymb,bm,graphicx}

\newcommand{\tou}[1]{{\boldsymbol{#1}}}
\newcommand{\tod}[1]{{\boldsymbol{#1}}}

\newcommand{\dd}{\tou{d}}
\newcommand{\ka}{\tou{\kappa}}
\newcommand{\A}{\tod{A}}
\newcommand{\D}{\tod{D}}

\def\og{\leavevmode\raise.3ex\hbox{$\scriptscriptstyle\langle\!\langle$~}}
\def\fg{\leavevmode\raise.3ex\hbox{~$\!\scriptscriptstyle\,\rangle\!\rangle$}}
\usepackage{color}
\usepackage{soul}
\newcommand{\corr}{\color{black}}
\newcommand{\correc}{\color{black}}
\begin{document}

{\title{ Variational  modeling adapted to the medium with gradient properties   }}


\author{
 Azdine Nait-ali\thanks{Institut Pprime -UPR CNRS 3346 - D\' epartement Physique et M\' ecanique des Mat\' eriaux ENSMA - T\' el\' eport 2, 1er avenue Cl\' ement Ader BP 40109 F86961 FUTUROSCOPE CHASSENEUIL Cedex FRANCE, azdine.nait-ali@ensma.fr, }, Sami Ben Elhaj Salah\thanks{sami-ben-elhaj-salah@ensma.fr}}
 

\maketitle
 
\begin{abstract}
This study aims to develop a numerical homogenization method that can be applied to a heterogeneous stratified medium. Traditional scale transition methods are inadequate in capturing the essential gradient properties of some materials. Therefore, the focus of this work is to construct a homogenized model that considers the material property gradient. To achieve this, a two-step homogenization scheme is proposed. Firstly, the 3D model is decomposed into multiple 2D heterogeneous layers, and the behavior of each layer is estimated using a micro-mechanical model such as the Hashin-Shtrikman bounds. Secondly, a variational sum method is used to rebuild the behavior of the 3D environment. Finally, the method is applied to homogenize a thin plate with a porosity gradient.

\end{abstract}

\section{ Introduction}
The gradient in material properties plays a significant role in various fields of application, such as aeronautics, electronics, and biomechanics. With the increasing use of additive manufacturing \cite{ Ng2023} or forming process \cite{ Chang2021} in these sectors, understanding the behavior of materials with property gradients has become more important. Several approaches have been proposed in the literature to account for property gradients, including non-local approaches \cite{eringen1972nonlocal, nait2017nonlocal, mindlin1968first, pham2011gradient, Aifantis, Francfort1999} and full-field simulations. However, these methods have limitations such as high computational costs and complexity.

In this study, we propose a numerical homogenization method to model the behavior of a heterogeneous stratified medium with property gradients. The method is based on the homogenization principle and consists of two stages. In the first stage, the 3D material is decomposed into multiple thin layers, and the behavior of each layer is estimated using the Hashin-Shtrikman bounds \cite{hashin1963variational}. In the second stage, the behavior of the 3D environment is reconstructed using a variational sum of 2D energy \cite{eringen1972nonlocal}.

The proposed method is applied to the case of a thin plate with a porosity gradient. The results demonstrate the effectiveness of the method in accurately capturing the behavior of materials with property gradients. This study provides insights into the behavior of materials with property gradients and offers a useful tool for engineers and researchers working in various fields of application.

\section{Variational sum}
The variational  method is proposed to reconstruct a 3D volumetric model from 2D surface models, based on the theory of plates. This method is suitable for materials or morphological parameters that vary continuously along a {\corr preferred }  direction, which is chosen to be the $x_3$ axis. A cube $\mathcal{O}$ with gradient properties along the $x_3$ axis is considered, and by {\corr rescaling}, the geometry is redefined as $\OO:=\hat{\OO} \times (0,1)$. The method aims to obtain an effective equivalent material that retains the properties that change according to $x_3$. The cube $\OO$ is divided into $n$ thinly layered plates, whose behavior is described by a law obtained by size reduction {\corr (3D to 2D). Indeed, a first homogenization is carried out by dimension reduction from a 3D plate with inclusions to a homogeneous 2D plate \cite{Michaille2014}. More specifically he resulting homogeneous 3D model is obtained by stacking the plates and applying a scale transition approach based on $\Gamma$-convergence. The proposed method takes into account the initial properties of the gradient. In some material cases with complex geometry, this division may be necessary, as shown in the example presented in \cite{Nait-Ali2014}}.

\begin{figure}[th]
\hspace*{0.2cm} 
\includegraphics[width=9cm,height=2cm]{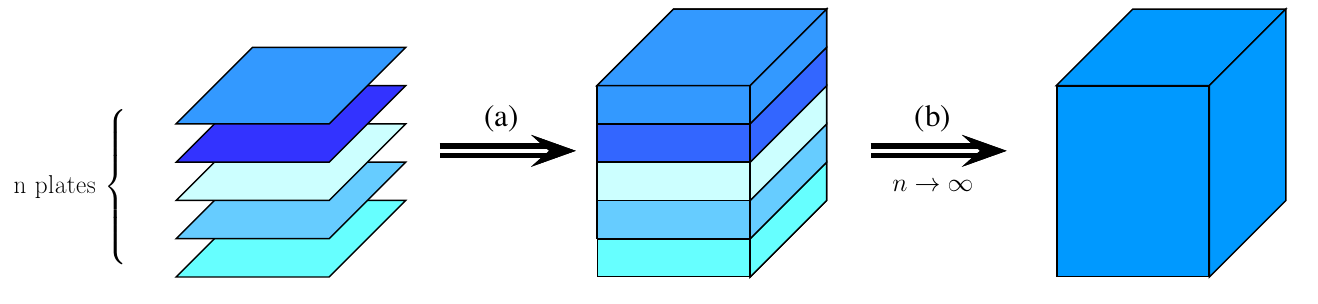}%
\caption{\corr Schema of the strategy of homogenization by variational summation.}
\label{fig2}
\end{figure}

the continuous energy can be obtained as the limit of the discrete energy as $n\rightarrow \infty$ and the thickness of the plates goes to zero. This leads to the following variational formulation:
\begin{eqnarray}
\inf_{u\in H^1(\OO;\R^3)} \int_{\OO} W(\defo u(x)) dx - \int_{\OO} L(x)\cdot u(x) dx
\label{energy-variational}
\end{eqnarray}
where $W$ is the stored energy density associated with the constitutive law of the material, and $L$ corresponds to an external load. The displacement field $u$ belongs to the Sobolev space $H^1(\OO;\R^3)$, which is the space of functions with square-integrable first derivatives. The function $\defo u(x)$ denotes the deformation gradient associated with the displacement field $u$, which describes the local stretching and rotation of the material at each point $x$ in the domain $\OO$.

{\corr Subfigure} (b) of Figure (1), the Riemann sum is transformed into an integral over the unit cube. The regularization energy term, denoted by $\gamma_n$, penalizes the variation of the energy between adjacent layers of the discretized model. It is defined as:

\begin{equation}
\gamma_n(u) = \frac{\lambda}{n^2}\int_{\OO} \sum_{i=0}^{n-2} \left(E_i(u) - E_{i+1}(u)\right)^2 d\mathbf{x}
\label{regularization-energy}
\end{equation}

where $\lambda$ is a positive regularization parameter. The regularization term tends to stabilize the energy by preventing large variations between adjacent plates. The total energy functional of the discrete model is then defined as:

\begin{equation}
\mathcal{E}_n(u) = E_n(u) + \gamma_n(u)
\label{total-energy}
\end{equation}
{\correc
$\gamma_n$ serves as the energy regularization function, which penalizes energy variation between adjacent layers of the discrete model. It's employed to stabilize the energy by preventing significant fluctuations between neighboring plates. Specifically, $\gamma_n(u)$ is defined by equation \ref{regularization-energy}, where $\lambda$ is a positive regularization parameter, and the summation over $i$ is taken across adjacent layers of the discretized model. And this function aims to approach $0$ as  $n$ tends to infinity, thereby achieving continuity of macroscopic energy.}

To summarize, the variational method presented in the work involves reconstructing a 3D volumetric model from 2D surface models using a size reduction technique. The resulting model takes into account the initial properties of the gradient {\corr as much as possible}. The energy associated with the set of plates is then minimized subject to appropriate boundary conditions and constraints, resulting in a homogenized material property. This homogenized property can then be used in the design and analysis of structures made of the original heterogeneous material.

{\correc
\begin{eqnarray}
 \widetilde \psi_0^n (x_3,s):=\psi_0(\frac{i}{n},s) \mbox{ and } 
 \widetilde L^n (x):=L(\hat x,\frac{ i}{n}) \\ \mbox{ if }\ x_3\in \textcolor{black}{[}\frac{i}{n}, \frac{i+1}{n}\textcolor{black}{[},
 \end{eqnarray}
This energy is the mechanical energy of each plate, which can be summed through integration, yielding  the energy  $E_n$ is given by:
\begin{eqnarray}
E_n( u)=
\dis \int_\OO  \widetilde \psi_0^n (x_3, \defo u) dx - \int_{ \OO} \widetilde L^n( x).u( x)dx\\
 \mbox{ if } u\in Step_{3,n}(\OO ) 
  \end{eqnarray}

The idea is therefore to make a two-step scale transition. We then define the average (or homogenized) energy density $\psi_0^n$, which is then a plate-type energy calculated on the average plane of the plate and for a representative volume element of $[0, m^2[$ with $m$ satisfying the following minimum condition.
\begin{equation}
 \psi_0^n(x_3, \defo u)  = \inf_{m\in \N^*}\{ \frac{1}{m^2}\int_{[0, m[^2 } S(x_3, \defo u)d\hat x\}
  \end{equation}
The function $S(.,.)$ is the energy density relative to the modeled linear behavior. For example, one can consider a linear elasticity model, but it can be extended to other physical behaviors. This function must be periodic within the study domain and possess additivity properties over the domain (i.e., summing over the entire domain yields the total behavior). 
Equation (8) corresponds to the average behavior over a representative elemental volume $[0, \bar{m}[^2$, where $\bar{m}$ satisfies the infimum of this equation.

Provided that the following assumptions are verified.\\

}

The energy density $s\mapsto \psi_0(.,s)$ must be convex and verifies the Lipschitz properties \cite{Nait-Ali2014}:

\begin{equation}
  |\psi_0(x_3, s)-\psi_0(x_3, s')| \leq\ell |s-s' |(1+|s\vert ^{p-1}+|s'|^{p-1}) \label{g2},
\end{equation}
For all couples $(s, s')\in \R^3\times \R^3$ and $\ell$  is a positive constant in varying steering $x_3$. As can be seen that  $\widetilde \psi_0^n$ satisfies the following growth condition for $ p>  1$ according to $  x_3$.
There are two positive constants $\alpha$ and $ \beta$ which do not depend on   $n$ and satisfying for all $s \in \R^3$ {the following equation:

\begin{equation}
  \alpha |s|^p \leq   \widetilde \psi_0^n (x_3, s) \leq \beta(1+ |s|^p). \label{g1}
\end{equation}

for all fixed $x_3\in (0,1)$. The global energy $ (E_n) _ {n \in \ N} $, $ \Gamma $-converges \cite{Braides2007} weakly to $E_0$ concerning the weak $L^p$ topology for $p$. 
\begin{equation}
E_0 (u) :=\int_\OO \psi_0 (x_3,\defo u) dx - \int_\OO L(x) .u(x) dx.
\label{som-var}
\end{equation}
The ergodicity hypothesis states that a system will eventually explore all of its possible states, and that the time average of any observable quantity of the system will be equal to its ensemble average. In the context of the present work, it means that the homogenized properties obtained from the variational limit hold true for all possible realizations of the heterogeneous material with the same statistical properties, provided that the system is ergodic.{\corr It is important to note that without this assumption, the macroscopic behavior obtained will deviate from the actual behavior.}

In other words, the variational limit is a valid approximation of the actual behavior of the heterogeneous material as long as the material satisfies the ergodicity hypothesis. This is an important assumption, as there may be cases where the material does not satisfy this hypothesis, and the homogenized properties obtained from the variational limit may not accurately represent the actual behavior of the material.

\section{Application: a case of a porosity gradient plate}

The equation presented describes the homogenized energy of the porous plate $\mathcal{O}$ with a continuous gradient of porosity. {\corr As described in \cite{nait2014volumic} The method involves a convex morphology of the pores.}The plate is considered as a bi-phase material with an isotropic matrix and a random distribution of voids inclusions that change according to the $x_3$ direction. The plate is divided into $n$ plates $\hat{\mathcal{O}}$ with the same thickness $\epsilon$. The behavior of each plate $\hat{\mathcal{O}}$ is described by a 2D potential consisting of the strains in the plane carried by the two directions $x_1, x_2$, due to membrane effects and bending, respectively, and the tensors of the behavior of the plate $\hat{\mathcal{O}}$ written for a 2D problem denoted by $\A_i$ and $\D_i$, respectively.

The homogenized energy is derived from the variational result and is denoted by $\psi_0$. The expression of $\psi_0$ is given by the integral over the volume of $\mathcal{O}$ of the sum of two terms. The first term is the product of the strain tensor $\dd\A(x_3)\dd$ and $1/2$, while the second term is the product of the bending tensor $\ka\D(x_3)\ka$ and $x_3^2$, where $\dd$ and $\ka$ are the strains in the plane carried by the two directions $x_1, x_2$, due to membrane effects and bending, respectively. The behavior of the plate $\mathcal{O}$ is characterized by the tensors $\A(x_3)$ and $\D(x_3)$, which depend on the porosity gradient along the $x_3$ direction. The integral is normalized by the volume of $\mathcal{O}$.

From a micromechanical point of view, the plate $ \mathcal{O} $ is a bi-phase with an isotropic matrix ( {\corr Young’s modulus $E$ and Poisson's ratio   $ \nu $)} and a random distribution of voids inclusions porosity of $ \theta(x_3) $. Considering the porosity gradient, the approach based on the variational sum is adopted to derive the actual behavior of the plate. The medium $ \mathcal{O} $ is divided into $ n $ in plates $ \hat {\mathcal{O}} $ with the same thickness $ \varepsilon$ which its behavior is described by the following 2D potential (limiting itself to a Kirchhof type plate):

\begin{equation}
\frac{1}{2}\dd A_k\dd+
\varepsilon^2\frac{1}{2}\ka\D_k\ka,
\end{equation}

With  $\dd$ and $\ka$ are the strains in the plane carried by the two directions $\tou {e} _1, \tou {e} _2 $, due to membrane effects and bending, respectively, whereas $ \A_K$ (resp. $ \D_K$) denote tensors of the behavior of the plate $k$ written for a 2D problem. {\correc

In the case of the limit passage described earlier, equation (12) reads:

\begin{equation}
 \psi_0^n(x_3, \defo u)  = \frac{1}{\vert \OO_n \vert}\int_{\mathcal{O}_n}\left(\frac{1}{2}\dd\A_n\left(x_3\right)\dd+
\frac{1}{n^2}\frac{1}{2}\ka\D_n\left(x_3\right)\ka\right) dV  \end{equation}
with $\OO_n$ the plate $n$.

We then need to work on the functions $ \underline{A}_n$ (respectively, $ \Delta_n$) for the limit passage.

For the transition from 3D to 2D, we let $\varepsilon$ tend to zero. We then obtain a homogeneous plate energy with $\A(x_3)$ and $\D(x_3)$ as the homogenized  material property for the plate at position $x_3$.}The macroscopic energy of the plate is homogenized from the variational result (\ref{som-var}). Its expression is given by:

\begin{center}
\begin{equation}
\displaystyle\psi_0=\frac{1}{\vert \OO \vert}\int_{\mathcal{O}}\left(\frac{1}{2}\dd\A\left(x_3\right)\dd+
x_3^2\frac{1}{2}\ka\D\left(x_3\right)\ka\right) dV
\label{plaque-homogene}
\end{equation}
\end{center}

Note here that the variational sum, resulting in (\ref{plaque-homogene}), allows us to find the same result as that obtained from the laminate theory.
The estimation of these properties is obtained by homogenization in two steps:

First, each fold like $i$ is a porous medium defined by (matrix $ (E, \nu) $ and $ porosity\ \theta_i $).
{ \correc
\begin{eqnarray}
  \A_i:=\int_{\frac{i}{n}}^{\frac{i+1}{n}} \mathbb C_i^{hom} dx_3 \\
  \D_i:=\int_{\frac{i}{n}}^{\frac{i+1}{n}}x_3^2 \mathbb C_i^{hom} dx_3 
 \end{eqnarray}

Then, the local properties $ \A_i $ and $ \D_i $ are estimated by performing a homogenization of the tensor $\mathbb{C}_i$ defined classically as: 

\begin{equation}
\mathbb{C}_i = \dfrac{E_i}{1-\nu_i^2}\begin{pmatrix}
1 & \nu_i & 0 \\
\nu_i & 1 & 0 \\
0 & 0 & \frac{1-\nu_i}{2}
\end{pmatrix}
\end{equation} 

 We then use variational methods of Hashin-Shtrikman to perform this homogenization, as well as the mixture law by computing averages on two elastic quantities of the same nature: $ \mathfrak{K} $ and $ \mu $ denote respectively the coefficients of compressibility and shear. They are linked to $ E $ and $ \nu $ by the following relationships:
}

\begin{equation}
{  \mathfrak  K} = \frac{E}{3(1-2\nu)}  \quad \mbox{and} \quad \mu=\frac{E}{2(1+\nu)}.
\end{equation} 
}

\begin{equation}
\frac{1}{ \mathfrak K^{HS}+\mathfrak K^*}=\frac{c^m}{ \mathfrak K^m+\mathfrak K^*}+\frac{c^f}{ \mathfrak K^f+\mathfrak K^*} \quad \mbox{with} \quad \mathfrak K^*= \mathfrak K^0
\end{equation} 
and 
\begin{equation}
\frac{1}{\mu^{HS}+\mu^*}=\frac{c^m}{\mu^m+\mu^*}+\frac{c^f}{\mu^f+\mu^*} \quad \mbox{with} \quad \mu^*=\frac{\mathfrak K^0\mu^0}{ \mathfrak K^0+2\mu^0}
\end{equation}
For, the calculation, we suppose $\mathfrak K^0=\mathfrak K^m$, $\mu^0=\mu^m$ and $\mathfrak K^0=\mathfrak K^i$, $\mu^0=\mu^i$ to estimate the lower and the upper bounds, respectively. ${c^m}$ and ${c^f}$ denote respectively the volume fraction for the matrix and fiber phase. More precisely since we are in the case of a porous material, the lower bound is equal to 0 and the upper bound returns the Mori-tanaka estimation. Then, we can determine the homogenized behavior of each plate by using the computation results of $ {E}^{{hom}} $ and $ {\nu}^{{hom}} $ {\correc and with (16) $\mathbb{C}_i^{hom}$} .

Second, the laminated plate of behavior is obtained by homogenizing the different folds according to equation (\ref{plaque-homogene}). Finally, the obtained results are compared to results obtained by a classical full-field method based on the Fast Fourier Transform (FFT).

\subsection{ Numerical simulations and results} 

The aim of this section is to compare the macroscopic behavior of our model with a classical homogenization method. A possible choice is to use an algorithm based on the Fourier Transform.

For the computation, we use the same algorithm presented \textcolor{black}{in \cite{Hemery2017}} and implemented an in-house FFT solver FoXtroT \cite{FoxFFT}. Then, the  classical elastic problem for a periodic material is described by the  following system.

\begin{equation}
	\left\lbrace
		\begin{aligned}
			{ div}(\sigma) 	&=0\\
			\sigma(x)	&={\mathbb{C}(x)}:\varepsilon(x)\\
			\varepsilon	&=\frac{1}{2} (\nabla u+^T\nabla u)\\
			<\varepsilon>	&=E \\
			u(x)	&=E.x + u^*(x)  \mbox{ with  } u^*(x)  \mbox{ periodic }
		\end{aligned}
	\right.
\label{sytem}
\end{equation}

$u(x),\varepsilon(x),\sigma(x)$ denote respectively the displacement, strain and stress fields whereas $E$ denotes the macroscopic strain. The stiffness tensor is noted by ${\mathbb{C}}(x)$ which is depends on the position of the vector $x$.  To solve the local problem, all the full-field simulations were performed with an FFT algorithm developed an in-house code FoxTrot \cite{Hemery2017, FoxFFT}.

The algorithm uses the original version proposed by Moulinec and Suquet \cite{Moulinec1998, Suquet95}. This method is formulated with the polarization tensor $\tau(x)=({\mathbb{C}}(x)- {\mathbb{C}}^0 ): \varepsilon(x)$ and consists in solving  the elastic problem given in system (\ref{sytem}). This iterative scheme with macroscopic strain and the Green operator  $\hat \Gamma_0$ are given in the literature \cite{Suquet95, tran2012micromechanics}. The main steps used in the algorithm are exposed in the following subsection:

The symbolic notations $FFT$ and $FFT^{-1}$ denote the Fast Fourier Transform and its inverse, respectively. The convergence test consists in checking the equilibrium with a strain-controlled method. In real space, the convergence criterion is defined: 
$||\varepsilon^{i+1} - \varepsilon^i ||_2 <10^{-4 } |\varepsilon^0 |$.\\

Where $||.||_2$ is the $L^2$ norm and $|.|$, called Euclidean norm of second-order tensor.

The choice to use an FFT calculation is based on two reasons. On the one hand, we aim to perform large calculations and then avoid the mesh part. On the other hand, we will only compare the elastic energies, that is why not necessary to have rich information at the local level. In this section, we will start by describing the process of generating the images that will be used as support in the simulations. Then, the results are presented and discussed

\subsubsection{Porous media generation}

\begin{figure}[th]

   \centering
	\includegraphics[scale=0.3]{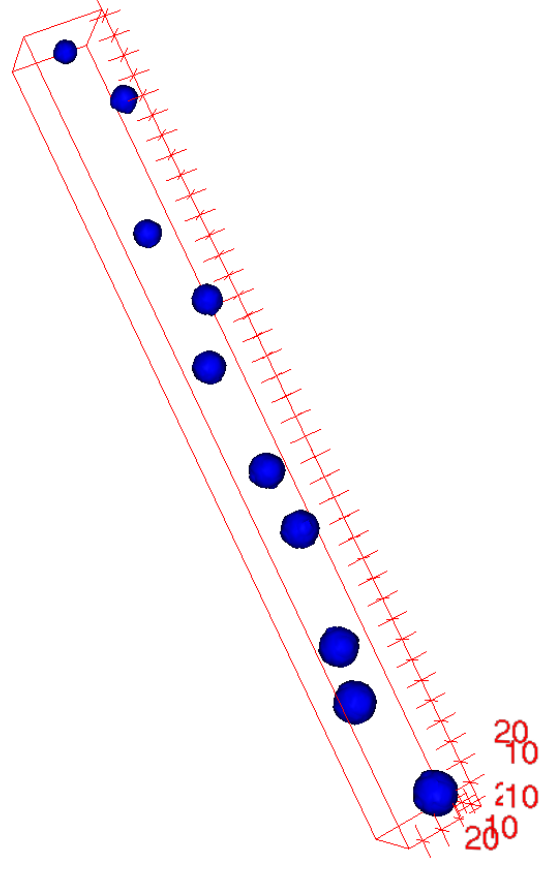}
		\caption{Unit cell 40x40x400 voxels using for generated bulk}

\end{figure}

\begin{figure}[th]
\centering
	\includegraphics[scale=0.3]{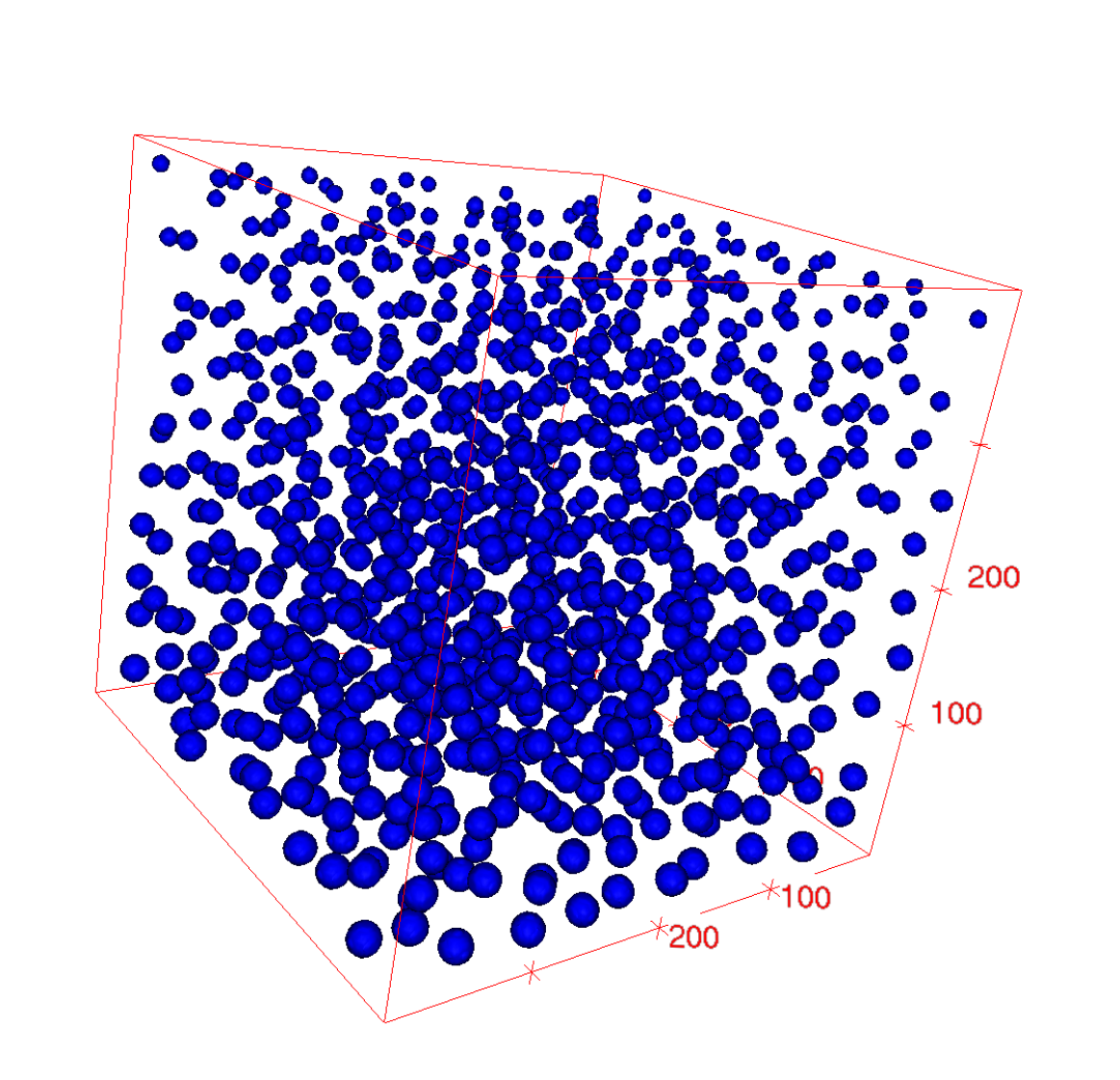}
		\caption{Porous media generated 400x400x400 voxels (2.2\% of porosity) }
\end{figure}

In the present work, the difficulty here is to generate a material containing spherical porosities of different sizes according to their position along the $e_{3}$-axis. Moreover, the modeling imposes as a hypothesis an ergodic distribution of the pores.For this purpose, a unit cell containing $n$ inclusion with varying radius along the $x_{3}$-axis is generated in a first cell (see {Figure (2)})
In order to guarantee the ergodicity hypothesis, we generated a random   $n\times n$ cells to obtain a cube/3D microstructure (see {Figure (3)})

\subsubsection{ Results }

{\corr When dealing with a material property gradient that varies continuously, outcomes derived from lamination and container theory are identified.  For all microstructure types and every inclusion volume fraction investigated, the Young  modulus of the inclusions was equal to 100 GPa while it was equal to 1 GPa for the matrix (contrast 100). The Poisson ratio for both constituents was 0.3}

 The variable $ \theta_i$ denotes the void ratio of plate $i$. We study 2 cases, the first is a deterministic one with a variation of porosity which depends linearly on the plate position. We can remark that the error measurements during the homogenization approach used in the present work are reasonable and still less than ($<10\%$).\\ 

For the interpretation of the results, we plot the variation of the global energy $\psi_0$ normalized by the same energy in the case without porosities, and this is according to the rate of porosities. The results can be explained by the fact that the variational sum described is an infimum bound. The same tendency of degradation of the coefficient of compressibility is also obtained for the homogenization by FFT.\\

$$
\theta_i:= i\theta_0.$$
\begin{figure}
	\centering
	\includegraphics[scale=0.3]{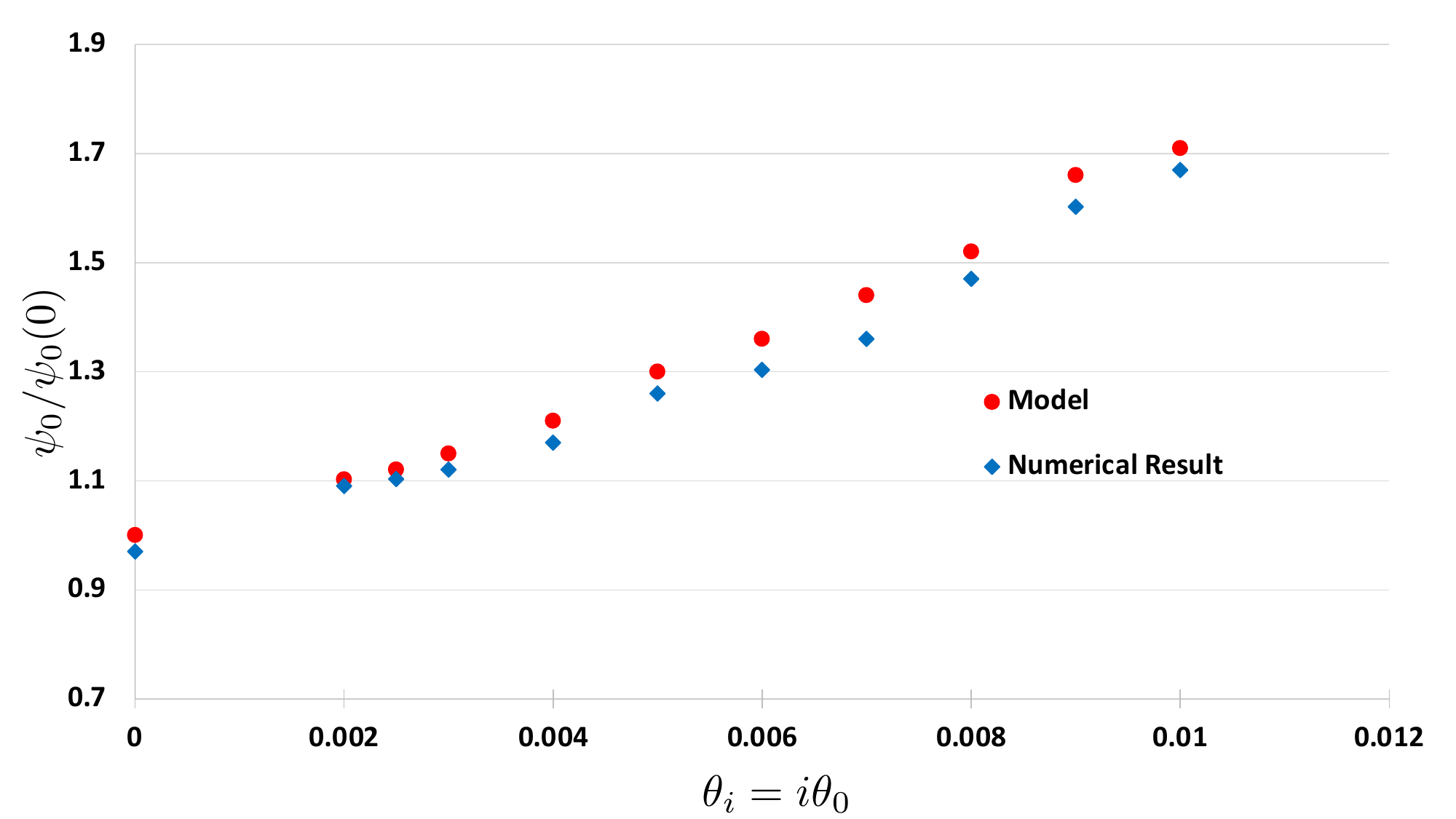}
	\caption{\corr Evolution of normalized energy density properties as a function of $\theta_i=i \theta_{0}$ in the linear case. The results from the proposed model are depicted in red, while those from the FFT simulation are shown in blue}
	\label{fig10}
\end{figure}

where $\theta_0$ denotes the initial porosity fraction for the plate $i$. The second case, is that the variation of porosity is quadratic:

$$\theta_i=i^2\theta_0$$

\begin{figure}
	\centering
	\includegraphics[scale=0.3]{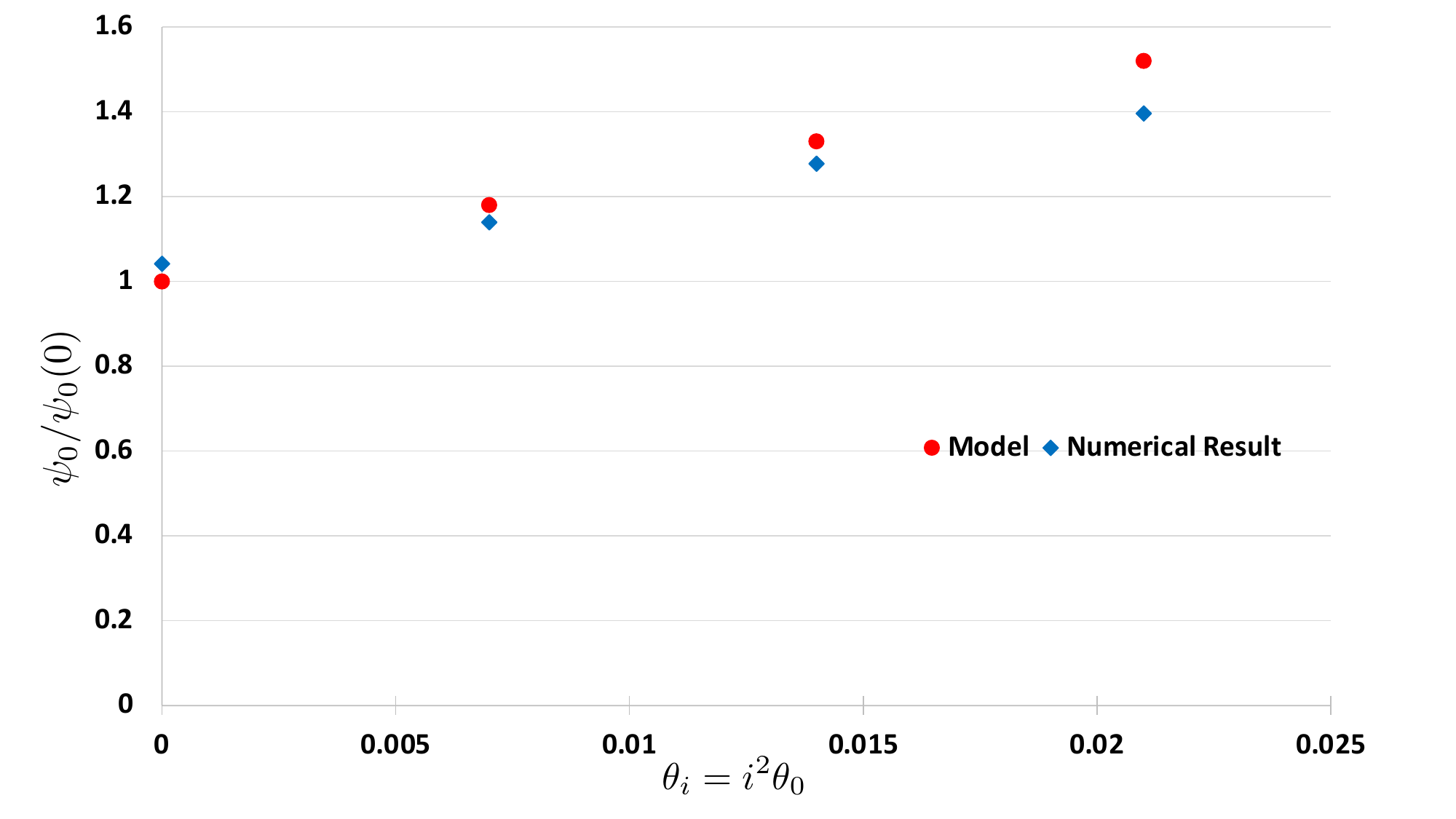}
	\caption{\corr Evolution of normalized energy density properties as a function of $\theta_i=i^2\theta_{0}$ in the quadratic case. The results from the proposed model are depicted in red, while those from the FFT simulation are shown in blue}
	\label{fig12}
\end{figure}

\newpage
{\correc To validate the model (Figure 6), we compared it with standard homogenization schemes. We show here the comparison with the Hashin-Shtrikman (HS) and Voigt (V) schemes. For these schemes, the discrepancy is significant because the calculation assumes a constant pore size and thus does not account for variations in size with $x_3 $. In the nonlinear case, it is observed that the simulation results deviate further from classical bounds, unlike our model. With a scheme that takes into account size variation, such as a self-coherent scheme, it could be more precise but would not consider the arrangement in plates. }

\begin{figure}
	\centering
	\includegraphics[scale=0.3]{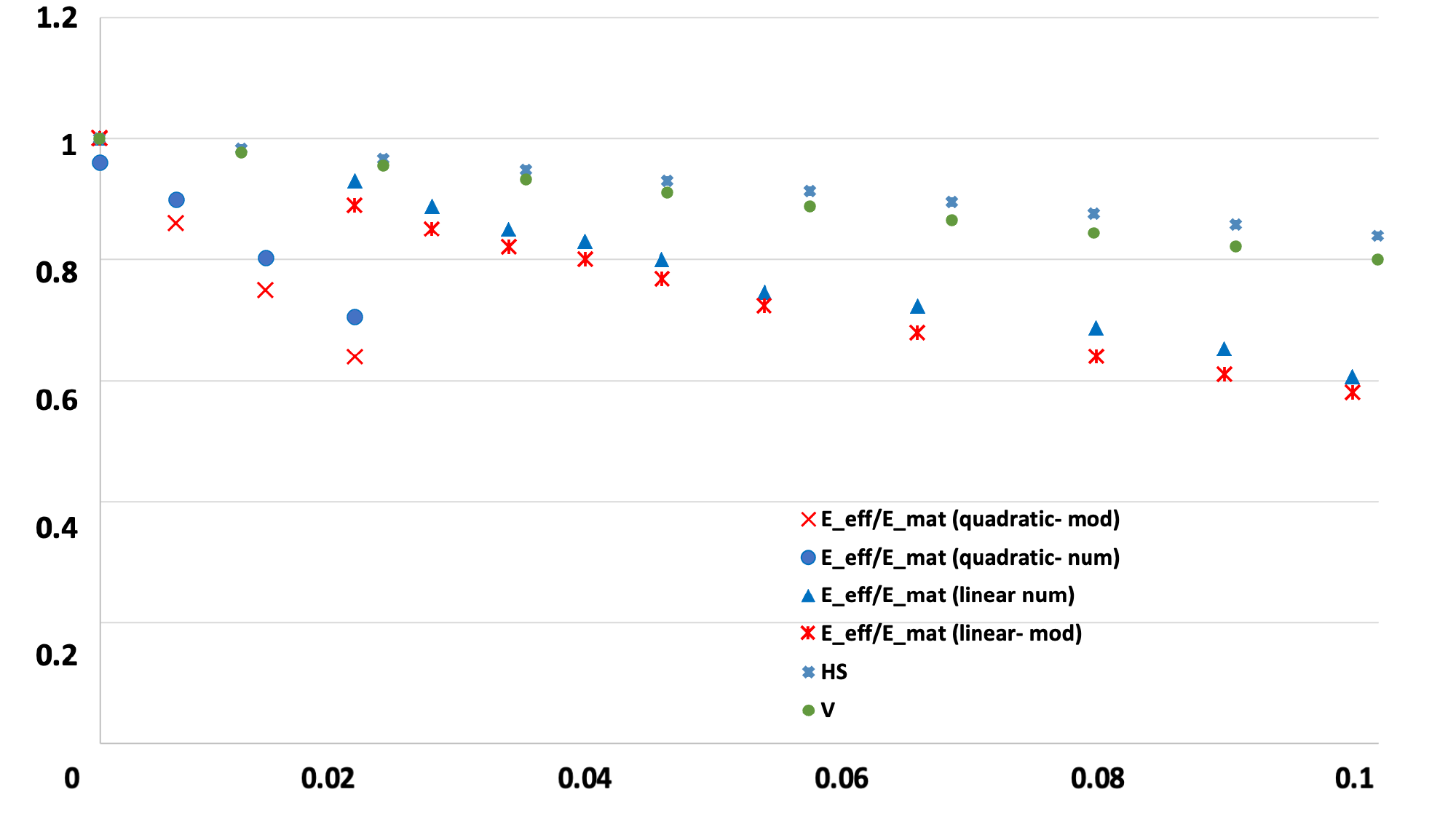}
	\caption{{\correc Evolution of normalized effective properties as a function of $\theta_i$ in quadratic and linear cases. In red are the results from the proposed model, in blue are the results from the FFT simulation. And comparison with the Hashin-Shtrikman (HS) and Voigt (V) }}
	\label{fig13}
\end{figure}

\newpage
\section{ Conclusions}

{\correc
In this article, the results demonstrate that the proposed model is in good agreement with the numerical homogenization performed in the linear case, presented in Figure (4) and Figures (6), as well as in the non-linear case, depicted in Figure (5) and Figures (6) {uncertain about including Fig.6}. This homogenization method allows us to naturally preserve the gradient properties (along the $x_{3}$ axis) using the parameter \(\theta\). Therefore, this model represents a good compromise between homogenization, which enables computations with low computational burden but entails information loss. Additionally, full-field models incur high computational costs (there is a 50-fold difference in CPU time between the computation time using FFT and our model for a 400x400x400 voxel case). For the simulation, we employ the internal solver \cite{FoxFFT}, which performs parallelized computation on 32 cores in 500 seconds CPU, whereas our script completes the computation in less than 10 seconds CPU.

Furthermore, in this study, we assume that within each plate, the pore distribution is ergodic  ( in there simulations  periodic with slight perturbation from the center of the inclusion), implying a homogeneous distribution. However, this method can be extended to cases of clusters \cite{test} or connected pores \cite{Gerard-Varet2022} distributed homogeneously. Therefore, the model must be adapted with appropriate homogenization between each plate. This approach could also be extended in future studies. Specifically, we will consider the parameter $\theta$ in a tensorial form to not only account for the unidirectional properties of the gradient.

}

\bibliographystyle{elsarticle-num}

\bibliography{MyCollection}

@article{Aifantis,
author = {Aifantis, K E and Willis, J R},
doi = {10.1016/j.mechmat.2005.06.010},
journal = {Mech. Mat.},
pages = {702--716},
title = {{Scale effects induced by strain-gradient plasticity and interfacial resistance in periodic and randomly heterogeneous media}},
volume = {38},
year = {2006}
}

@book{Braides2007,
author = {Braides, Andrea},
booktitle = {Gamma-Convergence for Beginners},
doi = {10.1093/acprof:oso/9780198507840.001.0001},
month = {sep},
publisher = {Oxford University Press},
title = {{Gamma-Convergence for Beginners}},
year = {2007}
}

@article{eringen1972nonlocal,
author = {Eringen, A Cemal},
journal = {Int. J. Eng. Sci.},
number = {1},
pages = {1--16},
publisher = {Elsevier},
title = {{Nonlocal polar elastic continua}},
volume = {10},
year = {1972}
}

@article{Francfort1999,
author = {Francfort, Gilles and Marigo, Jean-Jacques},
doi = {10.1051/proc:1999046},
issn = {1270-900X},
journal = {ESAIM Proc.},
pages = {57--74},
title = {{Une approche variationnelle de la m{\'{e}}canique du d{\'{e}}faut}},
volume = {6},
year = {1999}
}

@manual{FoxFFT,
author = {},
title = {{FoXTRoT: FFT-solver}},
url = {https://sourcesup.renater.fr/www/foxtrot/html/},
year = {2016}
}

@article{hashin1963variational,
author = {Hashin, Zvi and Shtrikman, Shmuel},
journal = {J. Mech. Phys. Solids},
number = {2},
pages = {127--140},
publisher = {Elsevier},
title = {{A variational approach to the theory of the elastic behaviour of multiphase materials}},
volume = {11},
year = {1963}
}

@article{Hemery2017,
author = {H{\'{e}}mery, S and Nait-Ali, A and Villechaise, P},
doi = {10.1016/j.mechmat.2017.03.013},
journal = {Mech. Mater.},
title = {{Combination of in-situ SEM tensile test and FFT-based crystal elasticity simulations of Ti-6Al-4V for an improved description of the onset of plastic slip}},
year = {2017}
}

@article{Michaille2014,
author = {Michaille, G{\'{e}}rard and Nait-Ali, Azdine and Pagano, St{\'{e}}phane},
doi = {https://doi.org/10.1093/amrx/abt007},
journal = {Appl. Math. Res. eXpress},
pages = {122--156},
title = {{Two dimensional deterministic model of a thin body with randomly distributed high conductivity fibers}},
volume = {1},
year = {2012}
}

@article{mindlin1968first,
author = {Mindlin, Raymond David and Eshel, N N},
journal = {Int. J. Solids Struct.},
number = {1},
pages = {109--124},
publisher = {Elsevier},
title = {{On first strain-gradient theories in linear elasticity}},
volume = {4},
year = {1968}
}

@inproceedings{Suquet95,
address = {Dordrecht},
author = {Moulinec, H and Suquet, P},
booktitle = {IUTAM Symp. Microstruct. Interact. Compos. Mater.},
editor = {Pyrz, R},
isbn = {978-94-011-0059-5},
pages = {235--246},
publisher = {Springer Netherlands},
title = {{A FFT-Based Numerical Method for Computing the Mechanical Properties of Composites from Images of their Microstructures}},
year = {1995}
}

@techreport{Moulinec1998,
author = {Moulinec, H and Suquet, P},
booktitle = {Comput. Methods Appl. Mech. Engrg},
institution = {lma},
isbn = {457825(97)00218},
pages = {69--94},
title = {{Computer methods in applied mechanics and englneerlng A numerical method for computing the overall response of nonlinear composites with complex microstructure}},
volume = {157},
year = {1998}
}

@article{Nait-Ali2014,
author = {Nait-Ali, Azdine},
doi = {10.1016/j.crme.2014.07.002},
issn = {16310721},
journal = {Comptes Rendus - Mec.},
title = {{Volumic method for the variational sum of a 2D discrete model}},
year = {2014}
}

@article{nait2014volumic,
author = {Nait-Ali, Azdine},
journal = {Comptes Rendus M{\'{e}}canique},
number = {12},
pages = {726--731},
publisher = {Elsevier},
title = {{Volumic method for the variational sum of a 2D discrete model}},
volume = {342},
year = {2014}
}

@article{nait2017nonlocal,
author = {Nait-ali, Azdine},
journal = {Comptes Rendus M{\'{e}}canique},
number = {3},
pages = {192--207},
publisher = {Elsevier},
title = {{Nonlocal modeling of a randomly distributed and aligned long-fiber composite material}},
volume = {345},
year = {2017}
}

@article{pham2011gradient,
author = {Pham, Kim and Amor, Hanen and Marigo, Jean-Jacques and Maurini, Corrado},
journal = {Int. J. Damage Mech.},
number = {4},
pages = {618--652},
publisher = {SAGE Publications Sage UK: London, England},
title = {{Gradient damage models and their use to approximate brittle fracture}},
volume = {20},
year = {2011}
}

@article{tran2012micromechanics,
author = {Tran, Thu-Huong and Monchiet, Vincent and Bonnet, Guy},
journal = {Int. J. Solids Struct.},
number = {5},
pages = {783--792},
publisher = {Elsevier},
title = {{A micromechanics-based approach for the derivation of constitutive elastic coefficients of strain-gradient media}},
volume = {49},
year = {2012}
}

@article{Ng2023,
author = {Ng, C. H. and Bermingham, M. J. and Dargusch, M. S.},
doi = {10.1016/j.addma.2023.103630},
issn = {22148604},
journal = {Additive Manufacturing},
number = {January},
title = {{Eliminating porosity defects, promoting equiaxed grains and improving the mechanical properties of additively manufactured Ti-22V-4Al with super-transus hot isostatic pressing}},
volume = {72},
year = {2023}
}

@article{Chang2021,
author = {Chang, Zhidong and Yang, Mei and Chen, Jun},
doi = {10.1016/j.jmatprotec.2020.117006},
issn = {09240136},
journal = {Journal of Materials Processing Technology},
number = {June 2020},
publisher = {Elsevier B.V.},
title = {{Experimental investigations on deformation characteristics in microstructure level during incremental forming of AA5052 sheet}},
volume = {291},
year = {2021}
}

@article{test,
author = {ElhajSalah, Sami and Nait-Ali, Azdine and Gueguen, Mikael and Nadot-Martin, Carole},
doi = {10.1016/j.jmatprotec.2020.117006},
issn = {09240136},
journal = {Mechanics of Materials},
number = {June 2020},
publisher = {Elsevier B.V.},
title = {{Asymptotic analysis,Heterogeneous material,Homogenization theory,Non-local phenomenon,Second gradient theory}},
volume = {147},
year = {2020}
}

@article{Gerard-Varet2022,
author = {G{\'{e}}rard-Varet, David and Girodroux-Lavigne, Alexandre},
doi = {10.3934/nhm.2022002},
issn = {1556181X},
journal = {Networks and Heterogeneous Media},
number = {2},
pages = {163--202},
title = {{Homogenization of Stiff Inclusions Through Network Approximation}},
volume = {17},
year = {2022}
}
\end{document}